\documentclass{PoS}

\usepackage[utf8]{inputenc}
\usepackage[T1]{fontenc}
\usepackage{slashed}
\usepackage{times}

\usepackage{amssymb,amsfonts,amsmath,amsthm}
\usepackage{dsfont,bbm}
\usepackage{dcolumn}

\usepackage{graphicx}
\usepackage[caption=false]{subfig}
\usepackage{color}

\newcommand{\MS}{{\overline{\mathrm{MS}}}}
\newcommand{\cO}{\mathcal {O}}
\newcommand{\Dl}[1]{\overset{\leftarrow}{D}_{#1}}
\newcommand{\Dr}[1]{\overset{\rightarrow}{D}_{#1}}
\newcommand{\Dd}[1]{\overset{\leftrightarrow}{D}_{#1}}

\title{Pion Distribution Amplitude from Lattice QCD}

\ShortTitle{Pion Distribution Amplitude}

\author{\speaker{V.M.~Braun}, S.~Collins, M.~G\"ockeler, P.~{P\'erez-Rubio}, A.~{Sch\"afer}, R.~W.~{Schiel}, A.~{Sternbeck} 
        \thanks{present address: Theoretisch-Physikalisches Institut,
     Friedrich-Schiller-Universit\"at Jena, 07743 Jena, Germany}\\
        Institut f\"ur Theoretische Physik, Universit\"at Regensburg, 93040 Regensburg, Germany\\
        E-mail: \email{vladimir.braun@ur.de}}

\abstract{We have calculated the second moment of the 
pion light-cone distribution amplitude using two flavors of 
dynamical (clover) fermions on lattices of different volumes, lattice spacings between 
$0.06 \, \mathrm {fm}$ and $0.08 \, \mathrm {fm}$ and 
pion masses down to $m_\pi\sim 150 \, \mathrm {MeV}$. 
Our result for the second Gegenbauer coefficient is  $a_2 = 0.1364(154)(145)$ 
and for the width parameter $\langle \xi^2 \rangle = 0.2361(41)(39)$. Both numbers refer to the 
scale $\mu=2 \, \mathrm {GeV}$in the $\MS$\ scheme, the first error 
is statistical including the uncertainty of the chiral extrapolation, 
and the second error is the estimated uncertainty coming from the 
nonperturbatively determined renormalization factors.}

\FullConference{QCD Evolution 2015 -QCDEV2015-\\
		26-30 May 2015\\
		Jefferson Lab (JLAB), Newport News Virginia, USA}

\begin{document}

\section{Introduction}

Hard exclusive processes involving energetic pions in the final state are 
sensitive to the momentum fraction distribution of the valence quarks 
at small transverse separations, usually called the pion distribution 
amplitude (DA). 
The DA is defined~\cite{Chernyak:1977as,Radyushkin:1977gp,Lepage:1979zb}
as the matrix element of a nonlocal light-ray quark-antiquark operator.
For example, for a positively charged pion
\begin{eqnarray}
\langle 0| \bar d(z_2n)\slashed{n}\gamma_5 [z_2n,z_1n] u(z_1n) |\pi(p)\rangle 
&=& if_\pi (p\cdot n) 
 \int_0^1 dx\, e^{-i(z_1 x + z_2 (1-x)) p\cdot n}\phi_{\pi}(x,\mu^2)\,, 
\label{def:phi}
\end{eqnarray}
where $p^\mu$ is the pion momentum, $n^\mu$ is a light-like vector, 
$n^2=0$, $z_{1,2}$ are real numbers, $[z_2n,z_1n]$ is the Wilson line 
connecting the quark and the antiquark fields and 
$f_\pi=132 \, \mathrm {MeV}$ is the 
usual pion decay constant. The DA $\phi_{\pi}(x,\mu^2)$ is scale-dependent, 
which is indicated by the argument $\mu^2$.
The definition in (\ref{def:phi}) implies the normalization condition
\begin{equation}
% \langle \xi^0 \rangle =   \int_0^1 dx\, \phi_{\pi}(x,\mu^2) = 1\,.
   \int_0^1 dx\, \phi_{\pi}(x,\mu^2) = 1\,.
\end{equation}

The physical interpretation of the variable $x$ is that the $u$-quark
carries the fraction $x$ of the pion momentum, so that $1-x$ is the
momentum fraction carried by the $\bar d$-antiquark.
Neglecting isospin breaking effects and electromagnetic corrections
the pion DA is symmetric under the interchange $x \leftrightarrow 1-x$:
\begin{equation}
   \phi_\pi(x,\mu^2) = \phi_\pi(1-x,\mu^2) \,.
\label{sym}
\end{equation} 
Moments of the DA, i.e. integrals with extra powers of the momentum fraction, 
are related to matrix elements of local operators and can be calculated on the lattice.
The symmetry in (\ref{sym}) implies that only the even moments involving the momentum fraction 
\emph{difference} 
%\begin{align}
$   \xi = x - (1-x) = 2x-1 $
%\end{align}
carry nontrivial physical information. Restricting to second order polynomials, usually one considers two 
types of moments: 
\begin{eqnarray}
 \langle \xi^2 \rangle(\mu^2) &=& \int_0^1 dx\, (2x-1)^2 \phi_\pi(x,\mu^2)\,, 
\nonumber\\
  \frac{18}{7} a_2 (\mu^2) = \langle C_2^{3/2}(\xi) \rangle(\mu^2) &=& \int_0^1 dx\, C_2^{3/2}(2x-1) \phi_\pi(x,\mu^2)\,, 
\end{eqnarray}
where $C^{3/2}_2(\xi) = (3/2)[5\xi^2-1]$ is a Gegenbauer polynomial. The advantage of using Gegenbauer moments is that 
the corresponding operators are multiplicatively renormalizable (in one loop) so that their matrix elements at
a certain reference scale are the true nonperturbative parameters that can be determined in lattice calculations.
The parameters $\langle \xi^2 \rangle(\mu^2)$ and $a_2 (\mu^2)$ in continuum theory are connected by a simple linear relation
\begin{align}
   a_2 = \frac{7}{12} \Big[ 5  \langle \xi^2 \rangle -1\Big]\,, 
&& 
  \langle \xi^2\rangle = \frac15 + \frac{12}{35} a_2\,,
\label{relations}
\end{align}
so that in principle they contain the same information. 
It is widely expected, however, that the numerical 
value of $\langle \xi^2\rangle$ is not far from 1/5 corresponding to the 
asymptotic pion DA   $\phi^{\mathrm {as}}_\pi(x) = 6 x(1-x)$ at $\mu^2\to \infty$. Hence, if
 $\langle \xi^2\rangle(\mu_0)$ is determined with a given accuracy at some reference scale 
$\mu_0$ by a certain nonperturbative method, and $a_2(\mu_0)$ is then obtained 
from the above relation, the error on $a_2$ is strongly 
amplified by the subtraction of the asymptotic contribution. 
The error on $a_2$ is the one that is relevant as it propagates through the 
renormalization group equations. In other words, although using 
$a_2$ as a nonperturbative parameter instead of $\langle \xi^2\rangle$ 
for the pion DA at a low reference scale $\phi_\pi(x,\mu^2_0)$ is just 
a rewriting, this choice is much more adequate in order to determine
the pion DA at high scales, $\phi_\pi(x,Q^2)$, $Q \gg \mu_0$, which 
enters QCD factorization theorems. Another issue to consider is that 
the relation in Eq.~(\ref{relations}) can be broken by lattice 
artifacts, see below.

\section{Lattice formulation}

We consider~\cite{Braun:2015axa} the following (bare) operators with two covariant derivatives 
as our operator basis:
\begin{align}
\cO^-_{\rho \mu \nu} (x)  
 & = \bar{d}(x) \left[ \Dl{(\mu} \Dl{\nu} 
  - 2 \Dl{(\mu} \Dr{\nu} + \Dr{(\mu} \Dr{\nu} \right] \gamma_{\rho)}  \gamma_5 \, u(x) \,, 
\nonumber \\
\cO^+_{\rho \mu \nu} (x)  
 & = \bar{d}(x) \left[ \Dl{(\mu} \Dl{\nu} 
  + 2 \Dl{(\mu} \Dr{\nu} 
  + \Dr{(\mu} \Dr{\nu} \right] \gamma_{\rho)} \gamma_5 \, u(x)\,. 
\label{eq:opmultiplets}
\end{align} 
Here $D_\mu$ is the covariant derivative and $(\ldots)$ denotes the 
symmetrization of all enclosed Lorentz indices and the subtraction of 
traces.  On the lattice the covariant derivatives will be
replaced by their discretized versions. 

The operator $\cO^-_{\rho \mu \nu}$ can be written in a conventional 
shorthand notation as
\begin{align}
\cO^-_{\rho \mu \nu} (x)  
 & = \bar{d}(x) \Dd{(\mu} \Dd{\nu} \gamma_{\rho)} \gamma_5 \, u(x) 
\end{align}
and its matrix element between the vacuum and the pion state is proportional 
to the bare lattice value of 
$\langle (x - (1 - x))^2 \rangle = \langle \xi^2\rangle$.
In the continuum, the operator 
$\cO^+_{\rho \mu \nu}$ is the second derivative of the axial-vector current:
\begin{equation} \label{eq:plusrelation}
\cO^+_{\rho \mu \nu} (x) = 
\partial_{(\mu} \partial_\nu \cO_{\rho)} (x) \quad \mbox{with} \quad
\cO_\rho (x) = \bar{d}(x) \gamma_{\rho} \gamma_5 u(x) \,.
\end{equation} 
Sandwiched between vacuum and the pion state, this relation 
guaranties energy conservation, i.e.  
 the bare value of $\langle (x + 1 - x)^2 \rangle = \langle 1^2 \rangle = 1$.
However, this identity is violated on the lattice because of 
discretization errors in the derivatives. 
The distinction between $\cO^+_{\rho \mu \nu}$ and 
$\partial_{(\mu} \partial_\nu \cO_{\rho)}$ for finite lattice spacing 
appears to be numerically important. 

The corresponding renormalized (e.g., in the $\MS$ scheme) 
axial-vector current is then given by
\begin{equation}
\cO^{\MS}_\rho (x) = Z_A \cO_\rho (x) 
\end{equation} 
with $Z_A \neq 1$ on the lattice. 
The operators $\cO^-_{\rho \mu \nu}$ and $\cO^+_{\rho \mu \nu}$
mix under renormalization even in the continuum. On the lattice the
continuous rotational $O(4)$ symmetry of Euclidean space is broken and 
reduced to the discrete $H(4)$ symmetry of the hypercubic lattice. 
This symmetry breaking can introduce additional mixing, in particular involving  
operators of lower dimension such that the mixing coefficients  
are proportional to powers of $1/a$, which complicates the 
renormalization procedure significantly.
The trick is  to choose the lattice operators such that they
belong to a particular irreducible representation of $H(4)$ that does not involve 
further operators, in particular of lower dimension. 
In the present case there is one such choice, corresponding to using 
operators $\cO^\pm_{\rho \mu \nu}$ with all three indices different.
In this calculation we restrict ourselves to the operators
(see, e.g., \cite{Braun:2006dg,Arthur:2010xf})
\begin{equation}\label{eq:operators}
  \mathcal O^{\pm}_{4jk}\,, \qquad j \neq k \in \{1,2,3\}\,.
\end{equation}
The renormalized operators are then given by
\begin{align}
\cO^{\MS -}_{4jk} (x) = Z_{11} \cO^-_{4jk} (x)
 + Z_{12} \cO^+_{4jk} (x) \,,
&&
\cO^{\MS +}_{4jk} (x) = Z_{22} \cO^+_{4jk} (x) \,.
\end{align} 
Note that due to the discretization artifacts in the derivatives
one cannot expect $Z_{22}$ to be equal to $Z_A$. 
The physical quantities of interest are given by matrix elements of the 
renormalized operators, e.g.  (in Euclidean notation)
\begin{align}
\langle 0 | \cO^{\MS}_4 (0) | \pi ({\bf p}) \rangle & = - i E_\pi ({\bf p}) f_\pi \,, 
\nonumber\\
\langle 0 | \cO^{\MS}_j (0) | \pi ({\bf p}) \rangle & = - p_j f_\pi \,.
\nonumber\\
\langle 0 | \cO^{\MS -}_{4jk} ( 0) | \pi ({\bf p}) \rangle 
& = i f_\pi \langle \xi^2 \rangle E_\pi ({\bf p}) p_j p_k \,,
\end{align} 
where  $p_j$ are the components of the three-vector $\bf{p}$ of the pion spatial momentum
%(defined as the contravariant space components of the Minkowski momentum $p_\mu$),
 and $E_\pi ({\bf p})$ is the corresponding energy.

The calculation of $\langle \xi^2 \rangle^{\MS}$ and $a_2^{\MS}$ involves 
two steps: computation of the bare matrix elements and
evaluation of the renormalization factors. We extract the bare matrix 
elements from two-point correlation functions of the operators 
$\cO^\pm_{\rho\mu\nu}$ and $\cO_\rho$ with suitable interpolating
fields $J(x)$ for the $\pi$-mesons. For the latter we consider the two 
possibilities
\begin{align}
 J_5(x) = \bar u(x)\gamma_5 d(x)\,,
&&
 J_{45}(x) = \bar u(x)\gamma_4\gamma_5 d(x) \label{eq:inter}
\end{align}
with smeared quark fields. The details of our smearing algorithm can be found in \cite{Braun:2015axa}.
Let
\begin{align}
 C^A_{\rho}(t,{\bf p}) &=  a^3 \sum_{\bf x} e^{-i{\bf p\cdot x}} 
\langle \cO_\rho({\bf x},t) J_A(0) \rangle,
\nonumber \\
C^{\pm;A}_{\rho\mu\nu}(t,{\bf p}) &= a^3 \sum_{\bf x}e^{-i{\bf p\cdot x}}
\langle \cO^\pm_{\rho\mu\nu}({\bf x},t) J_A(0) \rangle,
\label{eq:corr_function} 
\end{align}
where $A=5$ or $A=45$. The summation goes over the set of spatial lattice 
points ${\bf x}$ for a given Euclidean time $t$. 

For sufficiently large $t$, where the correlation functions are saturated by the 
contribution of the lowest-mass pion state, we expect that, e.g.,   
\begin{equation}
 C^{\pm;A}_{\rho\mu\nu}(t,{\bf p}) =
  \langle 0 | \cO^\pm_{\rho\mu\nu} (0)|\pi({\bf p})\rangle   
  \langle \pi({\bf p})|J_A(0) |0 \rangle \frac{1}{2E} 
  \left[e^{-Et} + \tau_{\mathcal O}\tau_{J}e^{-E(T - t)} \right] \,.
\end{equation}
Here $E \equiv E_\pi({\bf p})$, $T$ is the temporal extent of our 
lattice, and the $\tau$-factors take into account
transformation properties of the correlation functions under time reversal. 
One finds $\tau_{J_5} = -1$, $\tau_{J_{45}} = 1$, 
$\tau_{\cO} = 1$ for the operators $\cO^\pm_{4jk}$, $\cO_{4}$
and $\tau_{\cO} = -1$ for $\cO_j$, where $j,k=1,2,3$. 
We utilize these symmetries in order to reduce the statistical 
fluctuations of our raw data, i.e., we average over the
two corresponding times $t$ and $T-t$ with the appropriate sign factors.

From the ratios
\begin{equation} \label{eq:ratios-of-CFs}
 \mathcal R^{\pm;A}_{\rho\mu\nu; \sigma} =  
\frac{C^{\pm;A}_{\rho\mu\nu}(t,{\bf p})}{C^A_{\sigma}(t,{\bf p})}
\end{equation}
we can extract the required bare matrix elements 
$\langle 0 |\cO^\pm_{\rho\mu\nu} (0)|\pi({\bf p})\rangle$, which carry 
the information on the second moment of the pion DA.

A calculation of matrix elements of $\cO^\pm_{4jk}$ requires two nonvanishing spatial components 
of the momentum. We choose them as small as possible, $p = 2\pi/L$, 
where $L$ is the spatial extent of our lattice, and average over the possible directions, e.g., 
${\bf p} = (p,p,0)$, ${\bf p} = (p,-p,0)$, ${\bf p} = (-p,p,0)$, 
${\bf p} = (-p,-p,0)$ for $j=1$, $k=2$.
If the correlation functions are dominated by the single-pion states,
the time-dependent factors in the ratios of correlation functions 
cancel and we obtain, e.g., for the operator $\cO^\pm_{412}$ and 
the momentum ${\bf p} = (p,p,0)$
\begin{equation}
\mathcal R^{\pm;A}_{412; 4} = - \left(\frac{2\pi}{L}\right)^2 R^\pm \,, 
\end{equation}
where the constants $R^\pm$ are related to the bare 
lattice values of the second moment of the pion DA through
\begin{align}
  \langle \xi^2 \rangle^{\mathrm {bare}} = R^-,  &&
   a_2^{\mathrm {bare}} =  \frac{7}{12}\left( 5 R^- - R^+\right)\,.
\end{align}
They should not depend on the choice of the interpolating field $J_A$.
Note that $R^+ \slashed{=} 1$ and therefore for bare quantities
\begin{align}\label{eq:ineq}
    a_2^{\mathrm {bare}} &\slashed{=} \frac{7}{12}\left(5 \langle \xi^2 \rangle^{\mathrm {bare}} - 1\right)\,.
\end{align}
For the renormalized moments in the $\MS$ scheme we obtain
\begin{align}
 \langle \xi^2 \rangle^{\MS} = \zeta_{11} R^- 
+  \zeta_{12} R^+ \,,
&&
  a_2^{\MS} = \frac{7}{12} \Big[ 5 \zeta_{11} R^- 
  + \big(5 \zeta_{12} -  \zeta_{22}\big) R^+\Big] \,,\label{eq:ren_moments}
\end{align}
where 
\begin{equation} 
\zeta_{11} = \frac{Z_{11}}{Z_A}\,, \qquad
\zeta_{12} = \frac{Z_{12}}{Z_A} \,, \qquad
\zeta_{22} = \frac{Z_{22}}{Z_A}
\end{equation}
are ratios of renormalization constants defined in the next section.

In the continuum limit we expect that
\begin{align} 
Z_{22} \langle 0 | \cO^+_{4jk}(0) | \pi ({\bf p}) \rangle 
 &= - Z_A p_j p_k \langle 0 | \cO_4 (0) | \pi ({\bf p}) \rangle 
 \,=\, i p_j p_k E_\pi ({\bf p}) f_\pi \,.
\end{align}
Hence the quantity 
\begin{equation}\label{eq:I2} 
\langle 1^2 \rangle^{\MS} := \frac{Z_{22}}{Z_A}
\frac{\langle 0 | \cO^+_{4jk}(0) | \pi ({\bf p}) \rangle}
 {(-p_j p_k) \langle 0 | \cO_4 (0) | \pi ({\bf p}) \rangle } 
= \zeta_{22} R^+
\end{equation}
should approach unity as the lattice spacing tends to zero. In this case
the relation (\ref{relations})
%\begin{equation} 
%a_2^{\MS} = \frac{7}{12} \big( 5  \langle \xi^2 \rangle^{\MS} -1\big) 
%\label{a2-from-xi2}
%\end{equation} 
is recovered 
whereas for finite lattice spacing it follows from (\ref{eq:ren_moments})
\begin{equation} 
a_2^{\MS} = \frac{7}{12} \big( 5  \langle \xi^2 \rangle^{\MS} 
                             - \langle 1^2 \rangle^{\MS}\big)\,. 
\label{a2-from-xi2a}
\end{equation} 
E.g. for $a^2 \sim 5\cdot 10^{-3}$~fm$^2$ corresponding to 
$\beta = 5.29$, where most of our data are collected, we obtain
at $m_\pi = 294 \, \mathrm {MeV}$ on a $32^3 \times 64$-lattice
$
\langle 1^2 \rangle^{\MS}_{a \sim 0.07~\text{fm}} = 0.9402(66)(54) \,.
$
The deviation from unity is only 6\%, however, it results in a $25-30\%$
increase in the value of $a_2^\MS$ at the same lattice spacing, calculated 
using Eq.~(\ref{a2-from-xi2a}) instead of the continuum relation
in Eq.~(\ref{relations}). 
Approaching the continuum limit there are two possibilities: Either 
$\langle \xi^2\rangle$ is measured on the lattice, the result extrapolated 
to zero lattice spacing, and at the final step $a_2$ is obtained using 
the relation (\ref{relations}), or $a_2$ is calculated directly on the 
lattice and then extrapolated to the continuum limit. The first approach 
was used in Refs.~\cite{Braun:2006dg,Arthur:2010xf} whereas in this work 
we advocate using the second method.          

\section{Renormalization constants}

Bare matrix elements have to be renormalized and converted to the $\MS$ scheme which 
is used in the perturbative calculations. 
The first step is to choose lattice operators such that  
they belong to some irreducible, unitary 
representation of the symmetry group $H(4)$. In this way one can ensure that 
they do not mix with a larger class of operators, in particular with the operators
of lower dimension. Let  
$\cO^{(m)}_i (x)$ ($i=1,2,\ldots,d$, $m=1,2,\ldots,M$) be such a multiplet,   
see~\cite{Braun:2006dg,Arthur:2010xf,Braun:2015axa} for a concrete choice 
 Call the unrenormalized, but (lattice-)regularized
vertex functions (in the Landau gauge) $V^{(m)}_i (p,q)$, where $p$ 
and $q$ are the external quark momenta. The corresponding renormalized
(in the $\MS$ scheme) vertex functions are denoted by $\bar{V}^{(m)}_i (p,q)$.
The dependence of $\bar{V}^{(m)}_i$ on the renormalization scale $\mu$ 
is suppressed for brevity. Note that $V^{(m)}_i$ carries Dirac indices 
and is therefore to be considered as a $4 \times 4$-matrix. The color 
indices have been averaged over.

We choose a symmetric subtraction point
\begin{equation} 
p = \frac {\mu}{\sqrt{2}} (1,1,0,0) \; , \; 
q = \frac {\mu}{\sqrt{2}} (0,1,1,0) 
\end{equation}
such that $p^2 = q^2 = (p-q)^2 = \mu^2$. As our renormalization condition
we take (in the chiral limit)
\begin{eqnarray} 
\sum_{i=1}^d \mathrm {tr} 
    \left( \hat{B}^{(m)}_i \hat{B}^{(m') \dagger}_i \right)
 &=& Z_q^{-1}  \sum_{m''=1}^M \hat{Z}_{m m''} \sum_{i=1}^d 
   \mathrm {tr} \left( V^{(m'')}_i \hat{B}^{(m') \dagger}_i \right),
\end{eqnarray}
where $\hat{B}^{(m)}_i$ is the lattice Born term corresponding to $V^{(m)}_i$.
The wave function renormalization constant of the quark fields $Z_q$
is determined from the quark propagator, as usual~\cite{Gockeler:2010yr}, 
and subsequently converted to the $\MS$ scheme. Using the lattice Born 
term instead of the continuum Born term and proceeding analogously in the 
calculation of $Z_q$ ensures that $\hat{Z}$ is the unit matrix in the 
free case.

The renormalization matrix $\hat{Z}$ leads from the bare operators on
the lattice to renormalized operators in our SMOM scheme. The matrix
$Z$ transforming the bare operators into renormalized operators in the
$\MS$ scheme is then given by $Z = C \hat{Z}$, where the 
matrix $C$ is defined as
\begin{equation} 
\sum_{m''=1}^M \sum_{i=1}^d C_{m m''} \mathrm {tr} 
    \left( B^{(m'')}_i B^{(m') \dagger}_i \right)
 = \sum_{i=1}^d \mathrm {tr} 
   \left( \bar{V}^{(m)}_i B^{(m') \dagger}_i \right) .
\end{equation}
Here $\bar{V}^{(m)}_i$ is the renormalized vertex function in the
$\MS$ scheme and $B^{(m)}_i$ is the continuum Born term such that
the conversion matrix $C$ is completely determined from a continuum 
calculation~\cite{Gracey:2011fb,Gracey:2011zg}.

The calculation of the vertex functions with the help of momentum 
sources is straightforward. Partially twisted boundary conditions 
applied to the quark propagators allow us to vary the renormalization 
scale $\mu$ independently of the lattice size.  Due
to the rather small quark masses the subsequent chiral extrapolation 
appears to be quite safe.
  
The renormalization scale $\mu$ should, ideally, satisfy the conditions
$ 1/L^2 \ll \Lambda^2_{\mathrm {QCD}} \ll \mu^2 \ll 1/a^2 $
for a lattice with lattice spacing $a$ and extent $L$. 
In practice the $Z$-values at any given scale suffer from discretization artifacts 
as well as from truncation errors of the perturbative expansions. Therefore we try to 
exploit as much of the available nonperturbative information as possible
by performing a joint fit of the $\mu$-dependence of the chirally 
extrapolated renormalization matrices $Z(a,\mu)_{\mathrm {MC}}$ 
for our three $\beta$-values $\beta = 5.20$, $5.29$ and $5.40$. 
This is done using all available perturbative information and 
adding terms $\sim (a\mu)^{2k}$, $k=1,2,3$  as a plausible ansatz for an 
effective description of lattice artifacts.   

The statistical errors of the data are quite small, in particular for 
larger scales, so that the systematic uncertainties prove to be more important. 
The largest uncertainty comes from the variation of the number of loops in the 
conversion factors to the $\MS$\ scheme:
Working with the 1-loop vertex functions increases the result for
$\zeta_{11}$ by about 5\%, and the modulus of the mixing coefficient 
$\zeta_{12}$ increases even by about 17\%. We take two-loop results for the central values
and half of the difference between
the two-loop and the one-loop matching as the corresponding uncertainty. 
This should amount to a rather conservative error estimate.

In the previous paper~\cite{Braun:2006dg} the renormalization and mixing 
factors were evaluated in a mixed perturbative-nonperturbative approach, 
based on the representation of $\cO^+_{\rho \mu \nu}$ as the second 
derivative of the axial-vector current (see Eq.~(\ref{eq:plusrelation})).
Repeating this calculation in a completely nonperturbative setting
we find that the overall renormalization factor corresponding to 
$\zeta_{11}$ agrees within a few percent. The nonperturbative 
mixing coefficient, on the other hand, has the same (negative) sign 
as its perturbatively computed counterpart, but its modulus is up
to one order of magnitude larger. This observation underlines the 
necessity of nonperturbative renormalization, at least for the presently
reachable $\beta$-values.

\section{Results}

\begin{figure}[t]
\begin{center}
\includegraphics[width=7.0cm]{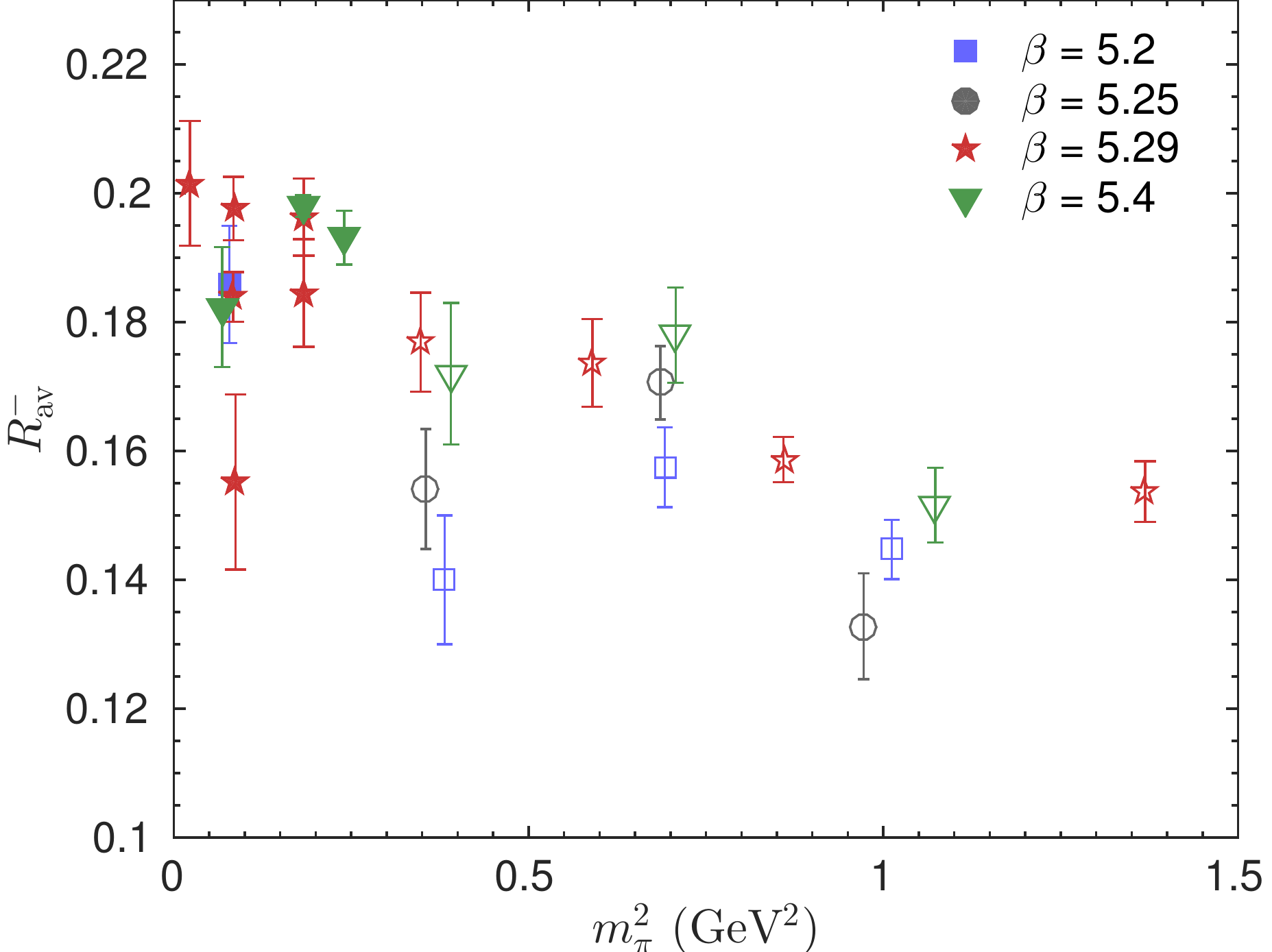}
\includegraphics[width=7.0cm]{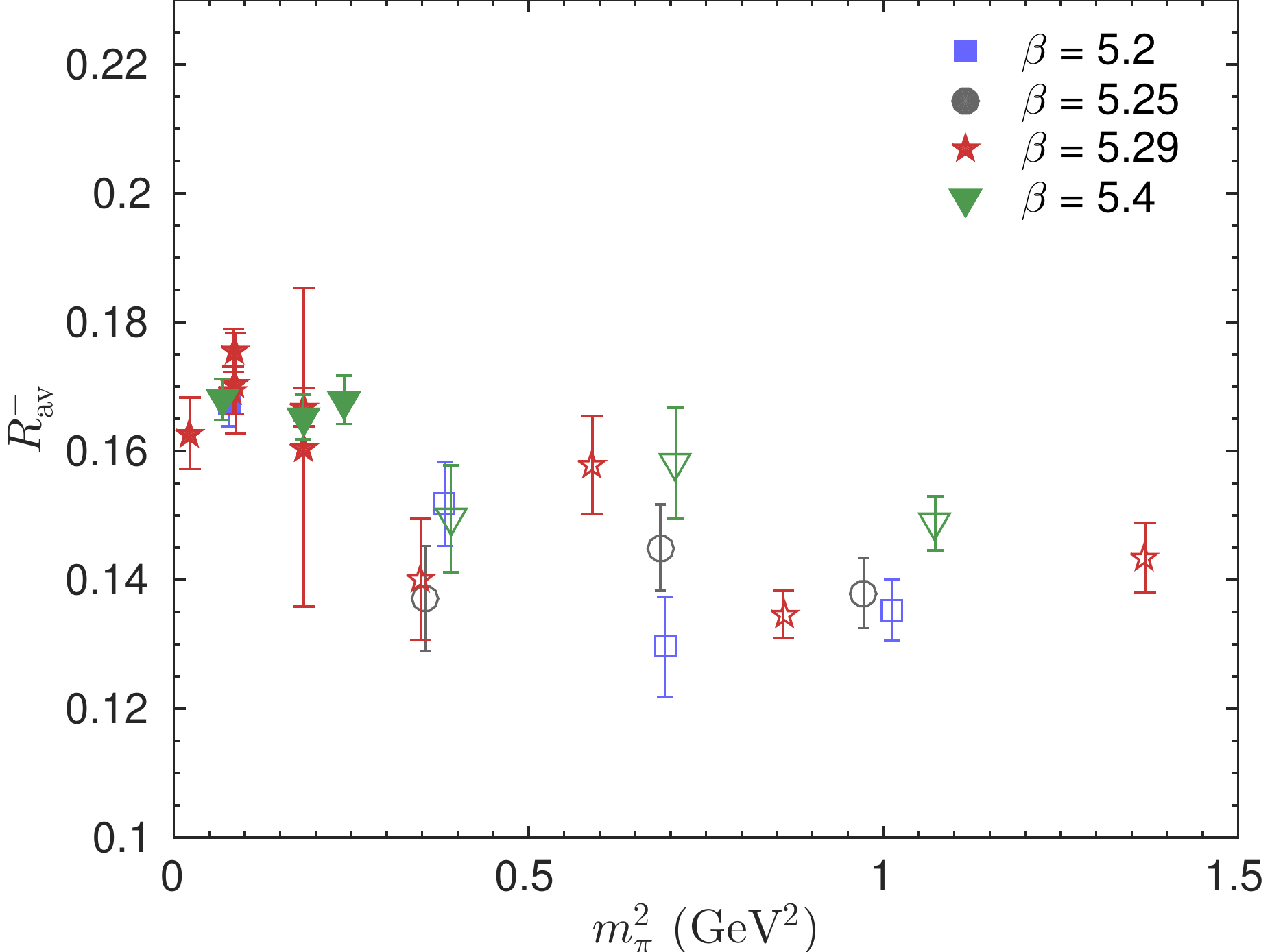}
\end{center}
\caption{\label{fig:all_data} Bare results for $R_{\mathrm {av}}^-$ 
from this work (filled symbols) and from \cite{Braun:2006dg} (open symbols)
for the two interpolators $J_{45}$ (left panel) and $J_5$ (right panel).}
\end{figure}

Bare lattice results for $R_{\mathrm {av}}^- = \langle \xi^2 \rangle^{\mathrm {bare}}$
for the two interpolating operators 
$J_{45}$ and $J_5$ are compared with the earlier study~\cite{Braun:2006dg} in
Fig.~\ref{fig:all_data}. Note that our data are consistent with the 
measurements in~\cite{Braun:2006dg}, but extend to
considerably smaller pion masses all the way down to the physical value.  
Nevertheless, we will see that taking into account Eq.~(\ref{eq:ineq}) and using 
the nonperturbatively computed value of mixing coefficient $\zeta_{12}$ leads to a 
significant shift in the final numbers.

\begin{figure}[t]
\begin{center}
\includegraphics[width=6.0cm]{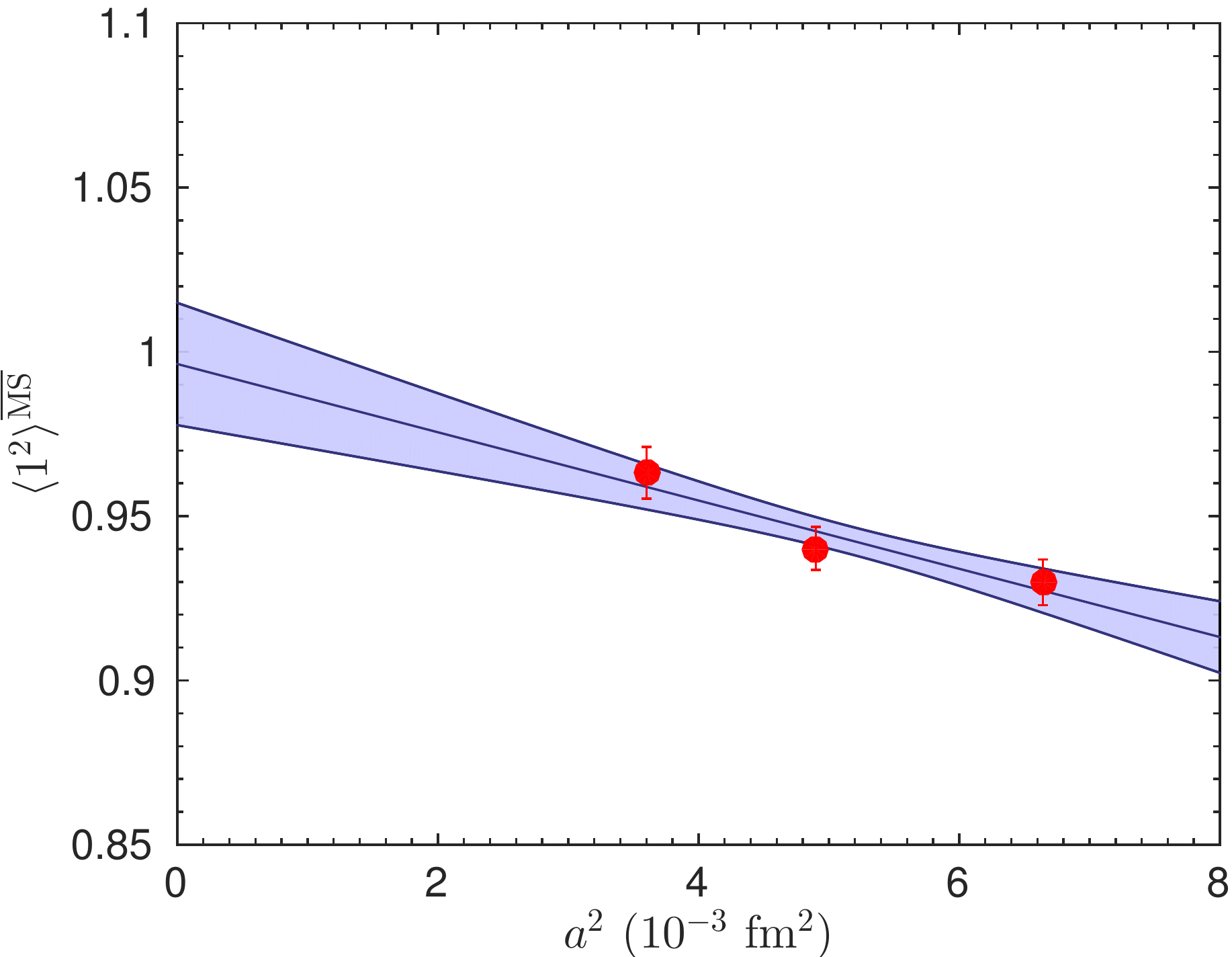}
\includegraphics[width=6.0cm]{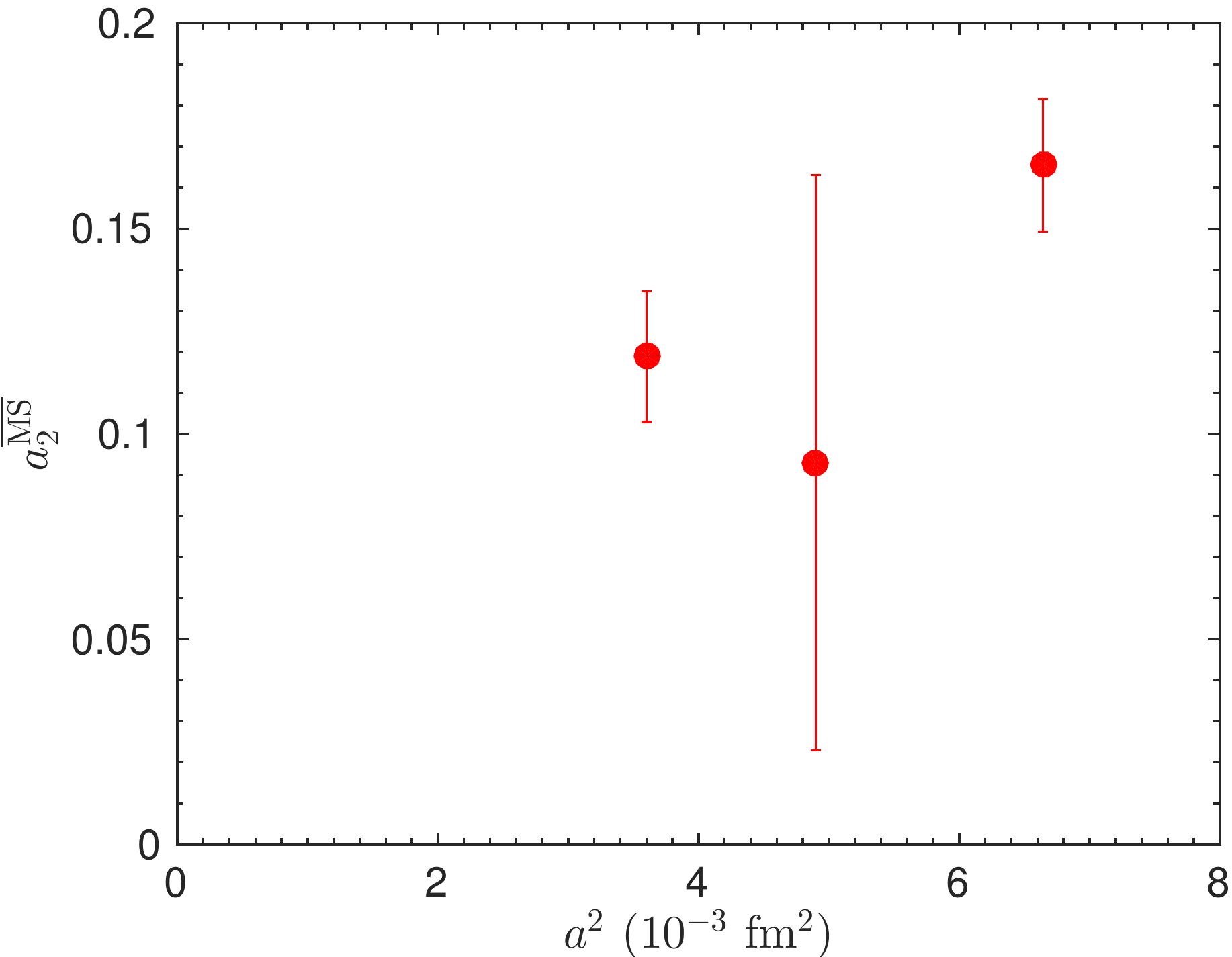}
\end{center}
\caption{ $\langle 1^2 \rangle^{\MS}$ [left panel] and $a_2^\MS$ [right panel] as a function of the lattice 
spacing $a$ for ensembles with $m_\pi L \sim 3.4 -3.8$ and 
$m_\pi \sim 280 \, \mathrm {MeV}$. Only statistical errors are shown.}
\label{fig:dscrete}
\end{figure}

The bare data are renormalized using the nonperturbatively computed
renormalization factors  $\zeta_{11}$, $\zeta_{12}$, $\zeta_{22}$, after which
the extrapolation to the physical pion mass, infinite volume, and eventually 
to the continuum has to be performed. The finite volume effects do not seem
to be significant so we do not discuss them here. The finite lattice spacing 
effects for  $\langle 1^2 \rangle^{\MS}$  and $a_2^\MS$ are illustrated in Fig.~\ref{fig:dscrete}.
For the former, the results can very nicely be extrapolated to the continuum limit $a=0$, 
reproducing the expected result. For the latter, however,  large statistical fluctuations do not
allow for an extrapolation. Although our data may show a tendency for $a_2^\MS$ (and $\langle \xi_2^\MS\rangle $) decreasing 
in the continuum limit, we do not consider this evidence as sufficient. By this reason we choose to present our final 
results for finite lattice spacing leaving the continuum extrapolation for 
a future study.  

\begin{figure}
\begin{center}
\includegraphics[width=7.0cm]{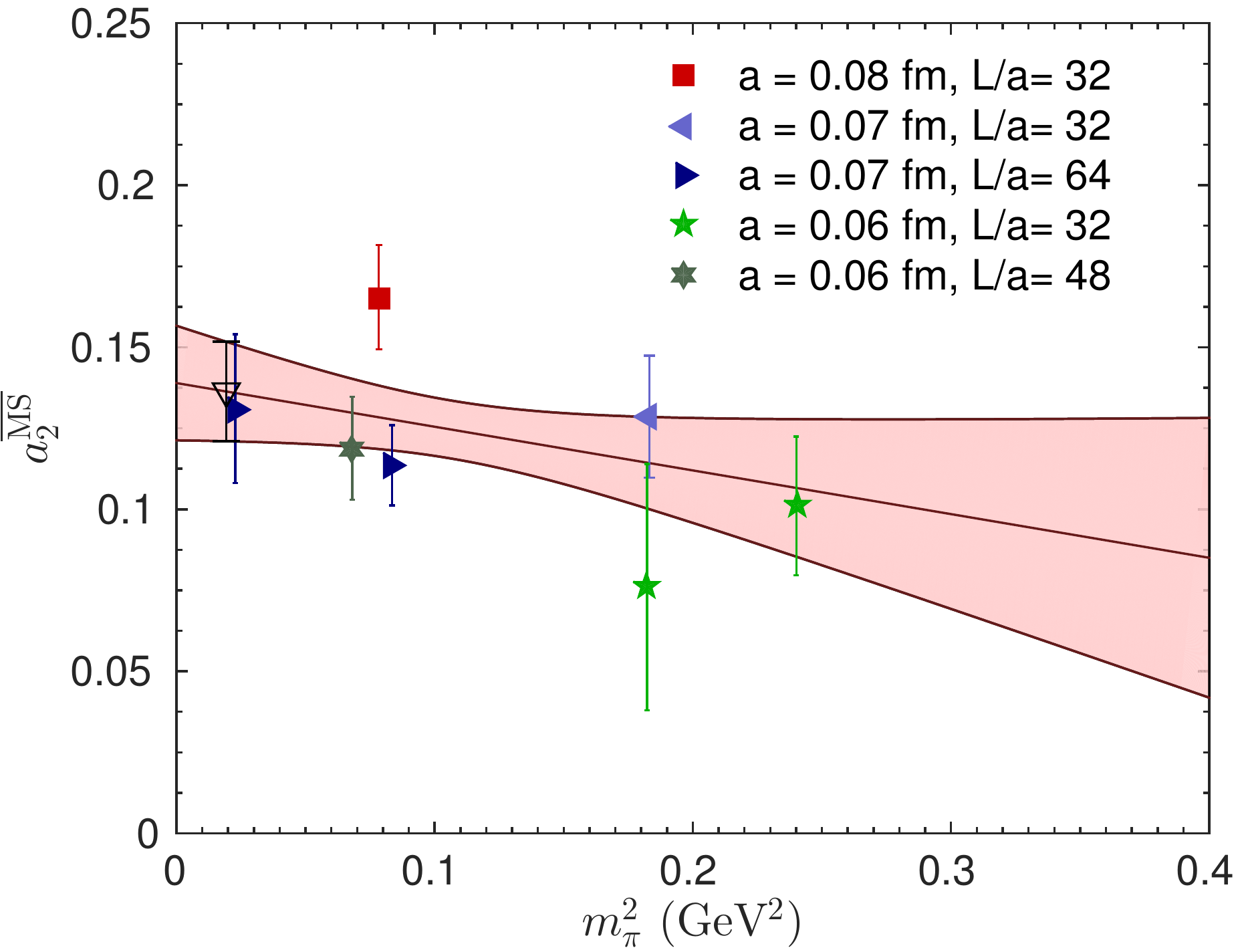}
\includegraphics[width=7.0cm]{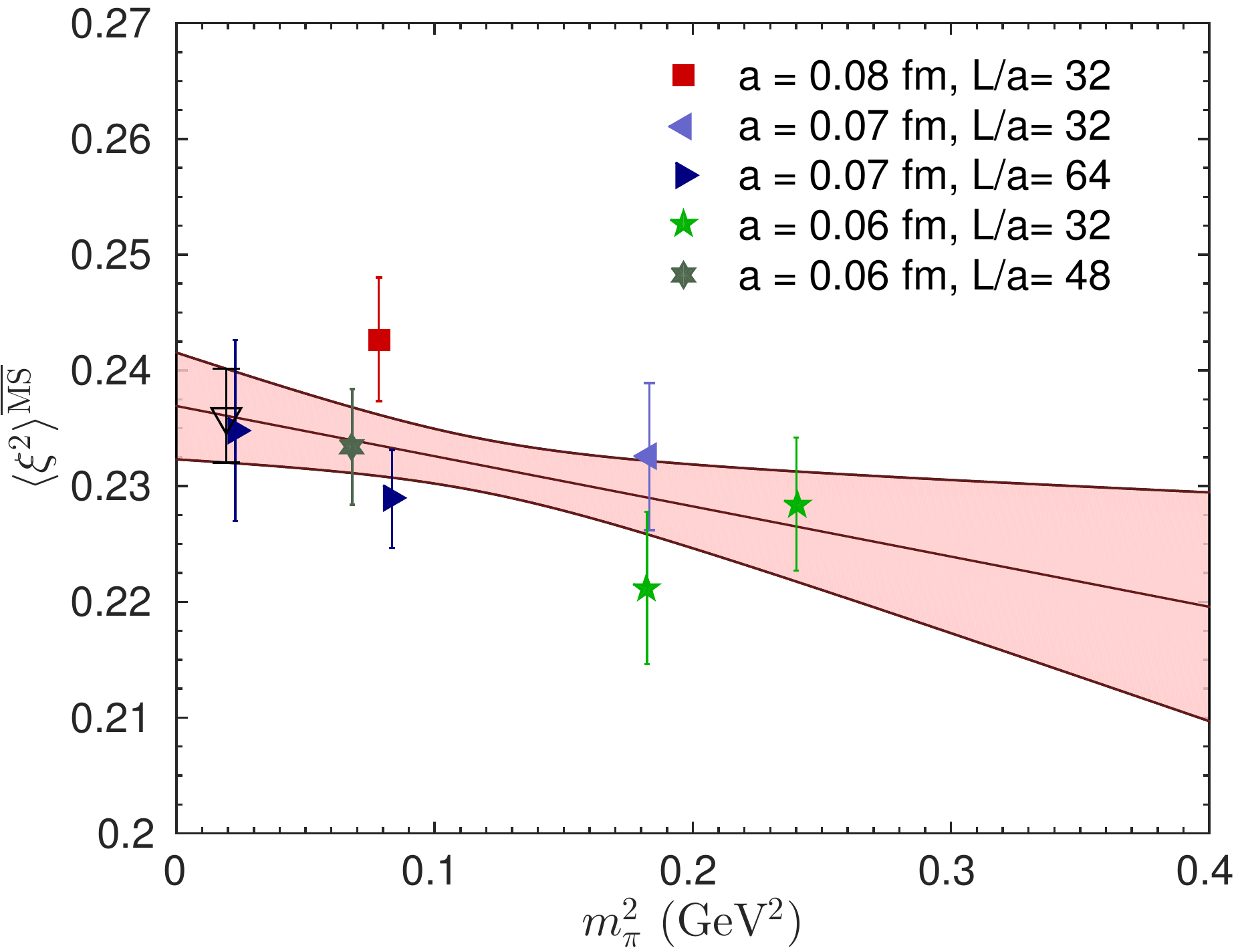}
\end{center}
\caption{\label{fig:chiral} Extrapolation to the physical
pion mass for $a_2^\MS$ [left panel] and $\langle\xi^2\rangle^\MS$ 
[right panel]. The open triangle represents the extrapolated value.
Only statistical errors are shown.}
\end{figure}

It is known~\cite{Chen:2005js} that $\langle \xi^2 \rangle^{\MS}$ and $a_2^{\MS}$ do not contain chiral logarithms, 
at least to one-loop order. Therefore we assume a linear dependence 
on $m_\pi^2$ for the extrapolation in the pion mass to the physical 
value. Since the ensemble with the lightest pion in our simulations is already 
very close to the physical point, the chiral extrapolation should be reliable.
As our lattice spacings do not vary that much,
and a proper continuum extrapolation of $\langle \xi^2 \rangle^{\MS}$ 
and $a_2^{\MS}$ cannot be attempted, we average the results from all 
lattice spacings but take into account only the data for the 
largest volume.
The resulting extrapolations of $a_2^{\MS}$ and $\langle \xi^2 \rangle^\MS$ to the physical pion mass are 
plotted in Fig.~\ref{fig:chiral}.

\section{Summary}

In this work we extend the lattice study~\cite{Braun:2006dg} of the 
second moment of the pion DA  by making use of a larger set of lattices 
with different volumes, lattice spacings and pion masses down to 
$m_\pi\sim 150 \, \mathrm {MeV}$ and implementing several 
technical improvements.
We employ the variational approach with two and three interpolators 
(not discussed above) to improve the signal from the pion state. 
The renormalization of the lattice data is performed nonperturbatively
utilizing a version of the RI'-SMOM scheme. For the first time we include
a nonperturbative calculation of the renormalization factor corresponding
to the mixing with total derivatives, which proves to have a significant 
effect. Our main result is 
\begin{equation}
 a_2 = 0.1364(154)(145)(?)
\label{result-a2}
\end{equation}
for the second Gegenbauer moment of the pion DA, and 
\begin{equation}
\langle \xi^2 \rangle = 0.2361(41)(39)(?) \,.
\label{result-xi2}
\end{equation}
They can be compared with the earlier lattice calculations
\begin{align}
    \langle \xi^2\rangle^\MS &=0.269(39)\,,\quad a_2^\MS = 0.201(114)\,, 
& \quad \text{\cite{Braun:2006dg}}
\nonumber\\
   \langle \xi^2\rangle^\MS  &=0.28(1)(2)\,,\quad a_2^\MS = 0.233(29)(58)\,. 
& \quad \text{\cite{Arthur:2010xf}}
\end{align}
All numbers refer to the scale $\mu = 2 \, \mathrm {GeV}$ in the $\MS$ scheme. 
The first error in (\ref{result-a2}) and  (\ref{result-xi2}) combines the statistical uncertainty and the 
uncertainty of the chiral extrapolation.
The second error is the estimated uncertainty contributed by the
nonperturbative determination of the renormalization and mixing
factors. Our lattice data are collected for the lattice spacing 
$a = 0.06-0.08 \, \mathrm {fm}$, and this range is not large enough to 
ensure a  reliable continuum extrapolation. 
The corresponding remaining uncertainty is indicated as (?). 
It has to be addressed in a future study.

As a final remark, we note that the somewhat 
smaller value of $a_2^\MS$ obtained in this work seems to be 
favored by the phenomenological studies of form factors in the 
framework of light-cone sum rules~\cite{Balitsky:1989ry}, 
see, e.g., Refs.~\cite{Agaev:2010aq,Agaev:2012tm,Ball:2004ye,Duplancic:2008ix,Khodjamirian:2011ub}.
Comparable numbers ($\langle \xi^2\rangle=0.25$, $a_2 = 0.15$) have also been obtained recently in the 
DSE approach in the calculation using DCSB-improved kernels~\cite{Chang:2013pq}.  

\acknowledgments

This work has been supported in part by the Deutsche Forschungsgemeinschaft 
(SFB/TR 55) and the European Union under the Grant Agreement IRG 256594.
We used gauge configurations generated by the QCDSF and RQCD collaborations. 
The computations were performed on the QPACE systems of the 
SFB/TR 55, Regensburg's Athene HPC cluster, the SuperMUC system at the 
LRZ/Germany and J\"ulich's JUGENE using the Chroma software 
system~\cite{Edwards:2004sx} and the BQCD software~\cite{Nakamura:2010qh}  
including improved inverters~\cite{Nobile:2010zz,LuscherOpenQCD}.
We thank John Gracey for helpful discussions about renormalization
issues and the UKQCD collaboration for giving us permission to use 
some of their gauge field configurations.

\end{document}